\documentclass{aa}
\usepackage{txfonts}
\usepackage{graphics}
\usepackage{natbib}
\usepackage{hyperref}

\newcommand{\be}{\begin{equation}}
\newcommand{\ee}{\end{equation}}
\newcommand{\beq}{\end{eqnarray}}
\newcommand{\eeq}{\end{eqnarray}}
\newcommand{\bc}{\end{center}}
\newcommand{\ec}{\end{center}}

\newcommand{\msun}{{M}_{\sun}}

\begin{document}

\title{Effects of Compton scattering on the neutron star radius constraints in rotation-powered millisecond pulsars}


\author{Tuomo~Salmi\inst{1}
\and Valery~F.~Suleimanov\inst{2,3,4}
\and  Juri~Poutanen\inst{1,4,5}}

\institute{Tuorla Observatory, Department of Physics and Astronomy, FI-20014 University of Turku, Finland 
    \and Institut  f\"ur  Astronomie  und  Astrophysik,  Kepler  Center  for  Astro  and  Particle  Physics,  Universit\"at  T\"ubingen,  Sand  1, 72076 T\"ubingen, Germany
    \and Astronomy Department, Kazan (Volga region) Federal University, Kremlyovskaya str. 18, 420008 Kazan, Russia
   \and Space Research Institute of the Russian Academy of Sciences, Profsoyuznaya str. 84/32, 117997 Moscow, Russia 
   \and Nordita, KTH Royal Institute of Technology and Stockholm University, Roslagstullsbacken 23, SE-10691 Stockholm, Sweden
}

\date{Received 11 March 2019 / Accepted 15 May 2019}

\abstract{
The aim of this work is to study the possible effects and biases on the radius constraints for rotation-powered millisecond pulsars when using Thomson approximation to describe electron scattering in the atmosphere models, instead of using exact formulation for Compton scattering. 
We compare the differences between the two models in the energy spectrum and angular distribution of the emitted radiation.
We also analyse a self-generated, synthetic, phase-resolved energy spectrum, based on Compton atmosphere and the most X-ray luminous,
rotation-powered millisecond pulsars observed by the Neutron star Interior Composition ExploreR (NICER).
We derive constraints for the neutron star parameters using both the Compton and Thomson models.
The results show that the method works by reproducing the correct parameters with the Compton model.
However, biases are found in both the size and the temperature of the emitting hotspot, 
when using the Thomson model.
The constraints on the radius are still not significantly changed, and therefore the Thomson model seems to be adequate if we are interested only in the radius measurements using NICER.
}

\keywords{stars: atmosphere --stars: neutron -- X-rays: binaries -- X-rays: stars}

\maketitle

\section{Introduction}
\label{sec:intro}

The equation of state of cold matter beyond nuclear densities can be constrained using astronomical observations of masses and radii of neutron stars (NSs) \citep{SLB10,lattimer12,NSK16,OF16,SPK16,WAC16,DS18,WYP19}.
In case of rapidly rotating NSs having radiating `hotspots' 
around their magnetic poles, we can model the observed pulses using general relativity and obtain constraints for their mass and the radius \citep{PG03,ML15}. 
However, detailed modelling requires knowledge of the spectral energy distribution and of the angular emission pattern of radiation emitted by the hotspots. 
The radiation escaping the hotspots is affected by energy-dependent absorption as well as by the anisotropic and energy-dependent scattering of photons by electrons in the atmospheres of NSs. 
 
There have been several studies aiming to constrain NS masses and radii using pulse profiles of accreting millisecond pulsars (AMPs), in which the matter from a low-mass companion star accretes onto the magnetic poles of the NS \citep[see e.g.][]{PG03,LMC08,P08,Ml11,SNP18}. 
However, these approaches suffer from a relatively high number of unknown NS parameters and from the uncertainties in the atmospheric structure, and therefore also in the angular and energy distribution of the emitted radiation.

In case of rotation-powered millisecond pulsars (RMPs), more independent information of the model parameters (e.g. mass and inclination) is often attained from radio data and the existing NS atmospheric models without effects of accretion may be used. 
In many RMPs, the bulk of X-ray radiation is thermal emission coming from the polar caps that are heated by a return flow of relativistic electrons and positrons in the open field line region \citep[see e.g.][]{HM02,bogdanov2018}. 
 Nevertheless, few RMPs exhibit nearly pure, non-thermal emission generated most probably by synchrotron emission from pulsar magnetospheres \citep{zavlin07}.
We focus on the thermally emitting RMPs, where the composition of the atmosphere is more confidently known than in AMPs (the RMP atmosphere likely consists of pure hydrogen instead of a mixture with heavier elements), and the temperature is low enough that the electron scattering presumably can be described using Thomson scattering approximation. 
The angular and energy distribution of the escaping photons can be described by using, for example, a plane-parallel atmosphere model in local thermodynamic equilibrium. 

 This type of model for RMPs that assumes Thomson scattering has previously been implemented in the McGill Planar Hydrogen Atmosphere Code (\textsc{McPHAC}), as described by  \citet{mcphac2012} \citep[see also e.g.][]{ZPS96,HRN06}. 
This code was also used by  \citet{miller2016} to simulate the data for RMP PSR J1614$-$2230 that can be provided by the Neutron star Interior Composition ExploreR (NICER), and to study the constraints on the NS mass and radius that can be obtained with those data. 
The question we ask in this paper is how an approximate treatment of Compton scattering affects the radiation spectra escaping from NS atmosphere and how this in turn affects the constraints on NS mass and radius from the NICER data. 
We note that exact treatment of Compton scattering is very important when considering NS atmospheres heated by accretion \citep{SPW18}, or by magnetospheric return currents as, for example, was recently discussed by \citet{BPO19} using a very simplified atmosphere model  \citep[see also e.g.][]{ZShak69,AW73,RS75,ZTZ95,GCZT19}. 
Modelling the heated RMP atmospheres is, however, beyond the scope of this work and will be discussed elsewhere.

The remainder of this paper is structured as follows. 
In Sect.~\ref{sec:methods}, we discuss the methods, including modelling the NS atmosphere, ray-tracing, and our method to create and analyse synthetic data. 
In Sect.~\ref{sec:results}, we first compare our spectral results to those computed with \textsc{McPHAC}, and then obtain NS parameter constraints fitting the data, which are created with the full Compton model, with both the full Compton and approximate Thomson scattering models.
We conclude in Sect.~\ref{sec:conclusions}. 

\section{Methods}\label{sec:methods}

We first constructed 
a model for NS atmosphere 
consisting of pure hydrogen. 
This is justified by the fact that, without the effects of continuing accretion, gravitational stratification leaves only the lightest elements in the atmospheric layers, which determine the properties of the escaping radiation.
We computed the atmosphere model and the angular distribution of the specific intensity of the escaping radiation using three different approaches.
In the first one we used our code  \citep{SPW12}, which treats Compton scattering using the exact relativistic Klein-Nishina cross-section and redistribution function derived and presented in details by \citet{AA81}, \citet{PKB86},  \citet{NP94a}, \citet{PouSve96}, and  \citet{PouVurm10}. 
As a second model, we used the same code, except 
where Compton scattering is treated in the Thomson limit.
This simplifies and accelerates the calculations dramatically. 
The third model was constructed using  \textsc{McPHAC} code, which also treats Compton scattering in the Thomson approximation (we used their anisotropic version of the model).

The parameters of the model are the effective temperature $T_{\mathrm{eff}}$ (which we will call just $T$ for brevity) and the surface gravity $g$. 
The solution of the equations that describe the NS atmosphere \citep[see e.g.][]{SPW12} provides us with the intensity of the escaping radiation. 
We tabulated these intensities over a grid of 360 photon energies, equally spaced in $\log E $ (keV) from $-$3.4 to 1.3, and 7 points in the cosine of the zenith angle $\mu$ (in the interval between zero and one using Gaussian nodes) for 11 values of temperature $T$ (K) spaced equally in $\log T$ from 5.5 to 6.6, and ten values of surface gravity $g$~({cm\,s}$^{-2}$) spaced equally in $\log g$ from 13.7 to 14.6.


The observed spectra depend on the NS mass, equatorial radius $R_{\mathrm{eq}},$ and spin (which determine gravitational acceleration $g$ as a function of co-latitude), and on the properties of the emitting spot, that is 
the local temperature $T$, the angular radius $\rho$, and the centroid (magnetic) co-latitude $\theta$.
The spectra also depend on the observer inclination (i.e. the angle between the line-of-sight and the NS rotation axis)
and the distance to the source. 
To compute the observed phase-resolved spectra and pulse profiles, we used `oblate Schwarzschild' approximation \citep[see e.g.][]{PB06,MLC07,ML15,SNP18}, taking into account the deformed shape of the star together with the special and general relativistic corrections to the photon trajectories and angles. 
For calculations of the total observed flux, integration over the spot surface is needed. 
However, in order to speed up the computations, the surface gravity $g$ for the atmospheric model was assumed to be constant within the spot (using the correct value for the spot centre). 
Thus, for each model we first needed only one piece-wise, two-dimensional, linear interpolation from the set of pre-calculated spectral tables to obtain a single two-dimensional array of intensities as functions of $E$ and $\mu$ only (corresponding to a given temperature $T$ and surface gravity $\log g$).
In our examples this is justified, because we considered only relatively small spots, where the changes in the NS radius within the spot are small.
Then, separately for each position within the spot, we again made a piece-wise, two-dimensional, linear interpolation to obtain intensities corresponding to a required photon emission zenith angle and energy.

The synthetic data were created keeping in mind the most promising NICER targets.
Instead of PSR J1614$-$2230 used by \citet{miller2016} as an example case, we focus on PSR J0437$-$4715 (the closest known RMP), or a similar pulsar with an expected high count rate (needed in order to observe any possible differences in the parameter constraints from the two spectral models). 
This pulsar has a complicated pulse profile presumably produced by two small high-temperature spots surrounded by a cooler annular region, and also an additional power-law component \citep{bogdanov2013}. 
However, since we aim only to compare the Thomson and Compton models, and are not necessarily interested in modelling this particular pulsar, we ignored these complications, and assumed two spots with constant temperature and pure thermal spectrum. 
PSR J0437$-$4715 was mainly used to obtain typical values for the parameters of the synthetic data. 
The model parameters are the following: spot temperature 
$T \approx$ 3.133 MK (0.27 keV), spot angular radius $\rho = 5.0^\circ$, spot co-latitude $\theta = 36 ^\circ$, equatorial radius of the star $R_{\mathrm{eq}} = 12$ km, and an arbitrary phase shift. 
These parameters were treated as free when fitting the data. 
Other model parameters were the NS mass $M=1.76~\msun$ 
, NS spin frequency $\nu = 173.6$ Hz, the distance to the star $D=156.3$ pc, the inclination $i = 42.4 ^\circ$, and neutral  hydrogen  column  density  for  interstellar  absorption $N_{\mathrm{H}} = 7 \times 10^{19} \mathrm{cm}^{-2}$ \citep[see e.g.][]{bogdanov2013, Deller2008, vbs08}.
They were regarded as fixed because they are or can be determined from other (radio) observations with relatively good accuracy.
We assumed that the observation of the source is long enough to accumulate the total number of observed counts of $4\times 10^{7}$. 

The fitting procedure of the data is mostly the same as presented in \citet{SNP18}.
We used Bayesian analysis and an affine invariant ensemble sampler \citep{goodman10} to obtain posterior probability distributions for the free model parameters.
The only exception is the phase shift, for which the maximum likelihood solution in each fit was found.
Additionally, the intrinsic scatter of the model was set as a free parameter $\log \sigma_{\mathrm{i}}$.
This is a measure of the systematic errors from the choice of the model 
\citep[see e.g.][]{SNP18}.
We assumed the prior probability distributions to be uniform in all of the parameters.
The limits of the priors were set to (11 km, 13 km) in $R_{\mathrm{eq}}$, (0$^\circ$, 90$^\circ$) in $\theta$, (1$^\circ$, 40$^\circ$) in $\rho$, 
(0.928 MK, 4.062 MK)  in $T$, and (0.868, 5.212) in $\log \sigma_{\mathrm{i}}$.
The synthetic pulse-profile data were binned into 16 phase bins and NICER energy channels located between 0.3 and 10 keV.
In addition, we required each modelled energy-phase bin to have more than 20 observed counts.

\begin{figure}
\resizebox{\hsize}{!}{\includegraphics{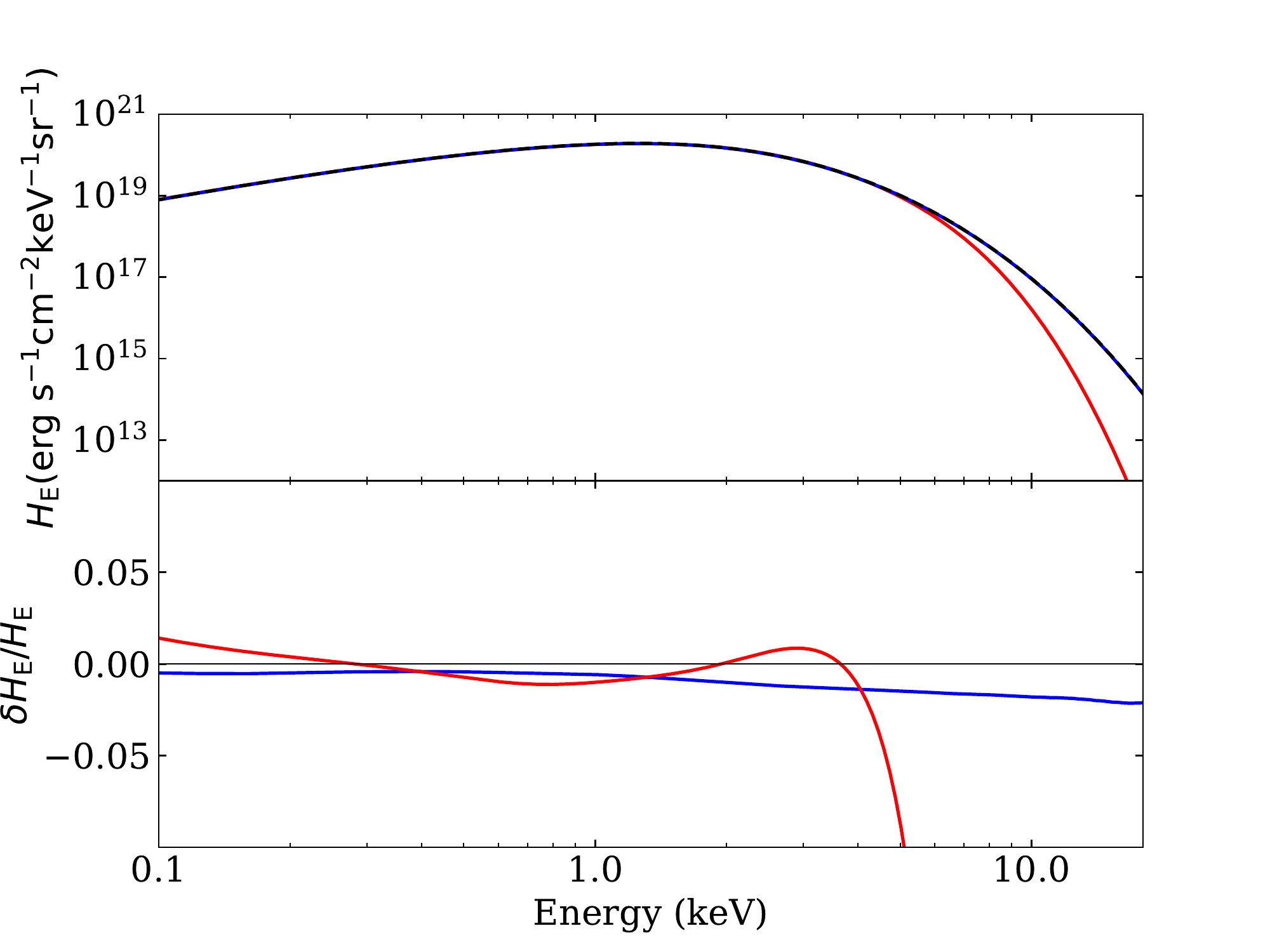}}
\caption{Upper panel: 
Model spectral energy distributions of first moment of specific intensity $H_{\mathrm{E}}$ for pure hydrogen NS atmosphere models with  
$T = $ 3.1623 MK and $\log g = 14.3856$. 
The outputs of the codes that use the Thomson approximation for Compton scattering are represented by the blue solid-line (our code) and black dashed-line (\textsc{McPHAC} code), while the red solid-line represents the output of our code when the full treatment of Compton scattering is used.
Lower panel: 
The relative difference between our model results in the Thomson (blue) and full Compton (red) limits compared to those of the \textsc{McPHAC} code.
}
\label{fig:thom_thom_spec}
\end{figure}

\begin{figure}
\resizebox{\hsize}{!}{\includegraphics{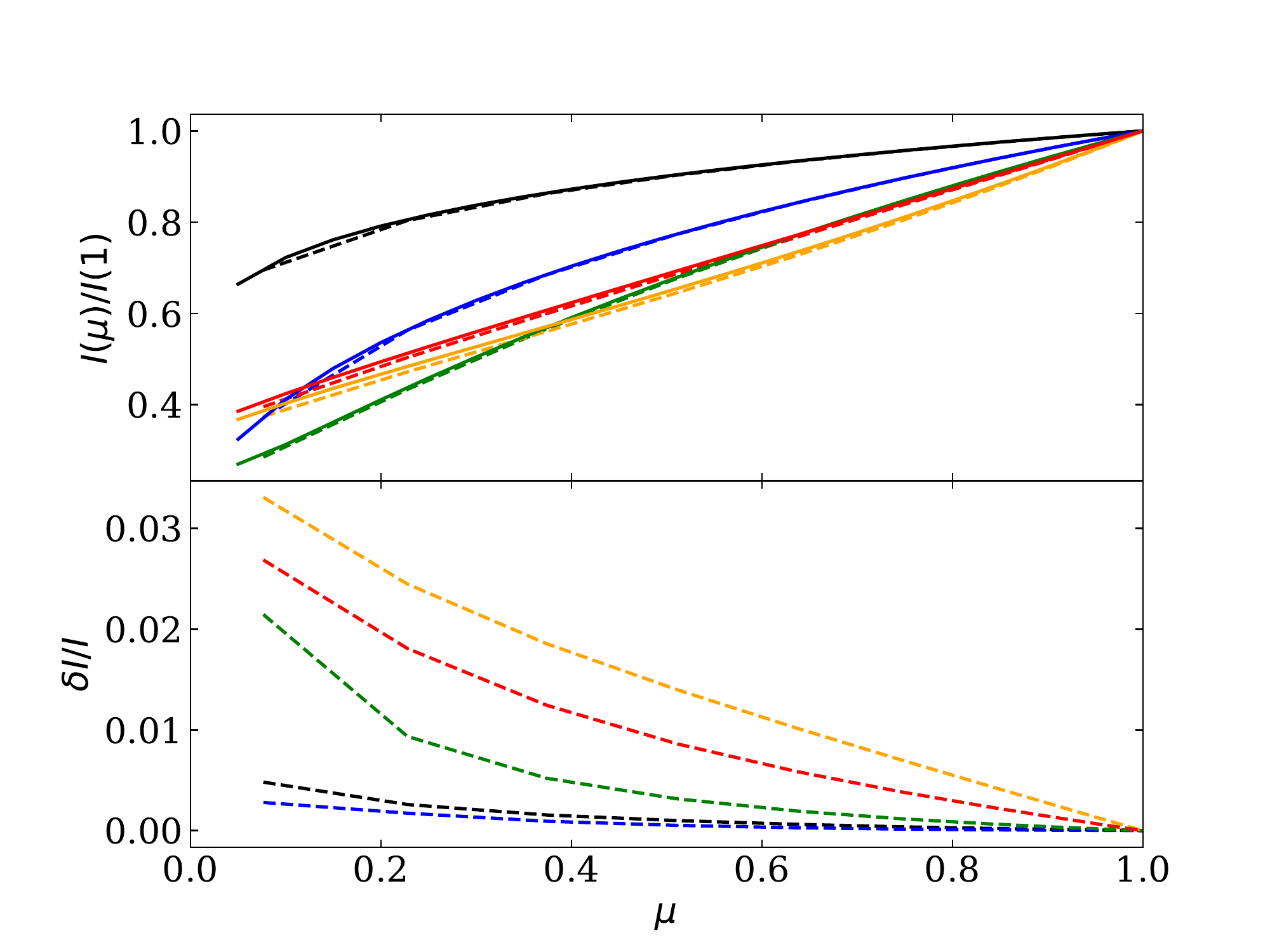}}
\caption{Upper panel: Angular distribution  of the specific intensity as a function of the cosine of the zenith angle $\mu$ for the NS atmosphere parameters given in Fig.~\ref{fig:thom_thom_spec}. 
The outputs of our code (using the Thomson approximation) and those of the \textsc{McPHAC} code are marked by solid- and dashed-lines, respectively.
The  black, blue, green, orange, and red colors correspond to 0.1,  0.5,  1.0, 5.0 , and 10.0 keV, respectively.
Lower panel: The relative difference between the normalised angular distributions is shown. 
}
\label{fig:thom_thom_angdeps}
\end{figure}

\section{Results}\label{sec:results}

\subsection{Spectral properties}

\begin{figure}
\resizebox{\hsize}{!}{\includegraphics{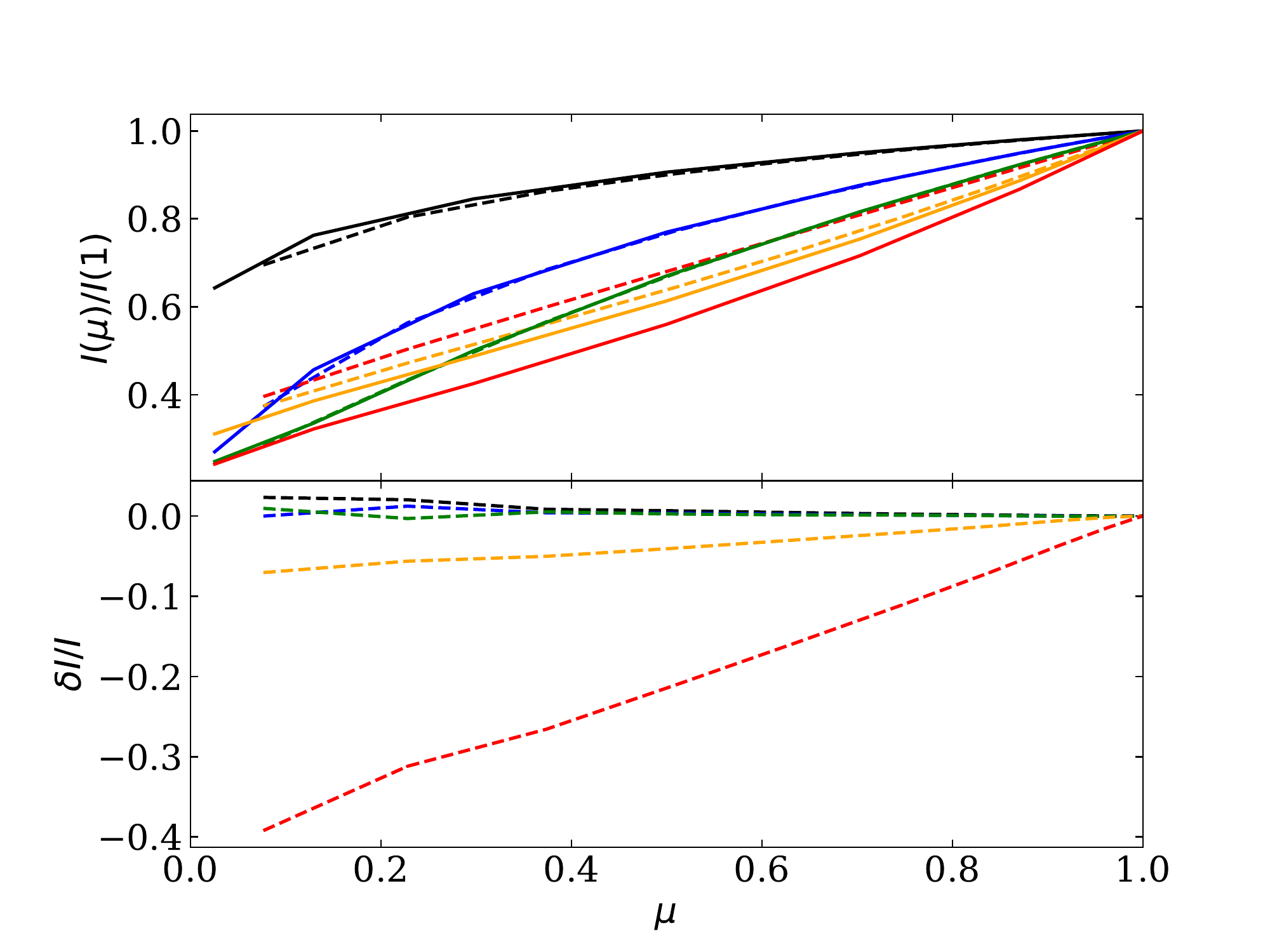}}
\caption{Comparison between Compton and Thomson models. Identifying information here is the same as in  
Fig. \ref{fig:thom_thom_angdeps}, but for full Compton scattering model (solid-lines) and Thomson model with \textsc{McPHAC} (dashed-lines).
}
\label{fig:thom_comp_angdeps}
\end{figure}

We began our calculations by checking that our code gives similar results to \textsc{McPHAC} when we used Thomson scattering instead of Compton. This is shown in Figs.~\ref{fig:thom_thom_spec} and \ref{fig:thom_thom_angdeps}, with the former showing the emergent spectrum and the latter the angular dependencies of the emitted radiation.
Figure~\ref{fig:thom_thom_spec} also shows the results computed with the full Compton model.
The comparison between the angular dependencies given by that model and those of \textsc{McPHAC} are shown in Fig. \ref{fig:thom_comp_angdeps}.
The parameters of the model, temperature and surface gravity, were chosen to be 
$T=$ 3.1623 MK and $\log g = 14.3856$, which are reasonable for RMPs. 

From the aforementioned figures, we see that the calculations with \textsc{McPHAC} agree 
with the Thomson version of our code within a few per cent, albeit displaying a small systematic discrepancy that increases with energy, and is probably connected to the increasing error at high zenith angles (i.e. small $\mu$) seen in Fig. \ref{fig:thom_thom_angdeps}.
The largest difference is about 3\%.
In addition, the effective temperature produced by \textsc{McPHAC} code is slightly higher, meaning the energy conservation is not extremely accurate. 
However, this should only have a minor effect to the fitted effective temperature.
A much larger difference is seen between our Compton and the Thomson models, which also becomes  more significant at higher energies (above 3 keV for the chosen temperature) and small $\mu$. 
This difference in spectrum is similar to that presented in \citet{SW07}.

\begin{figure}
\resizebox{\hsize}{!}{\includegraphics{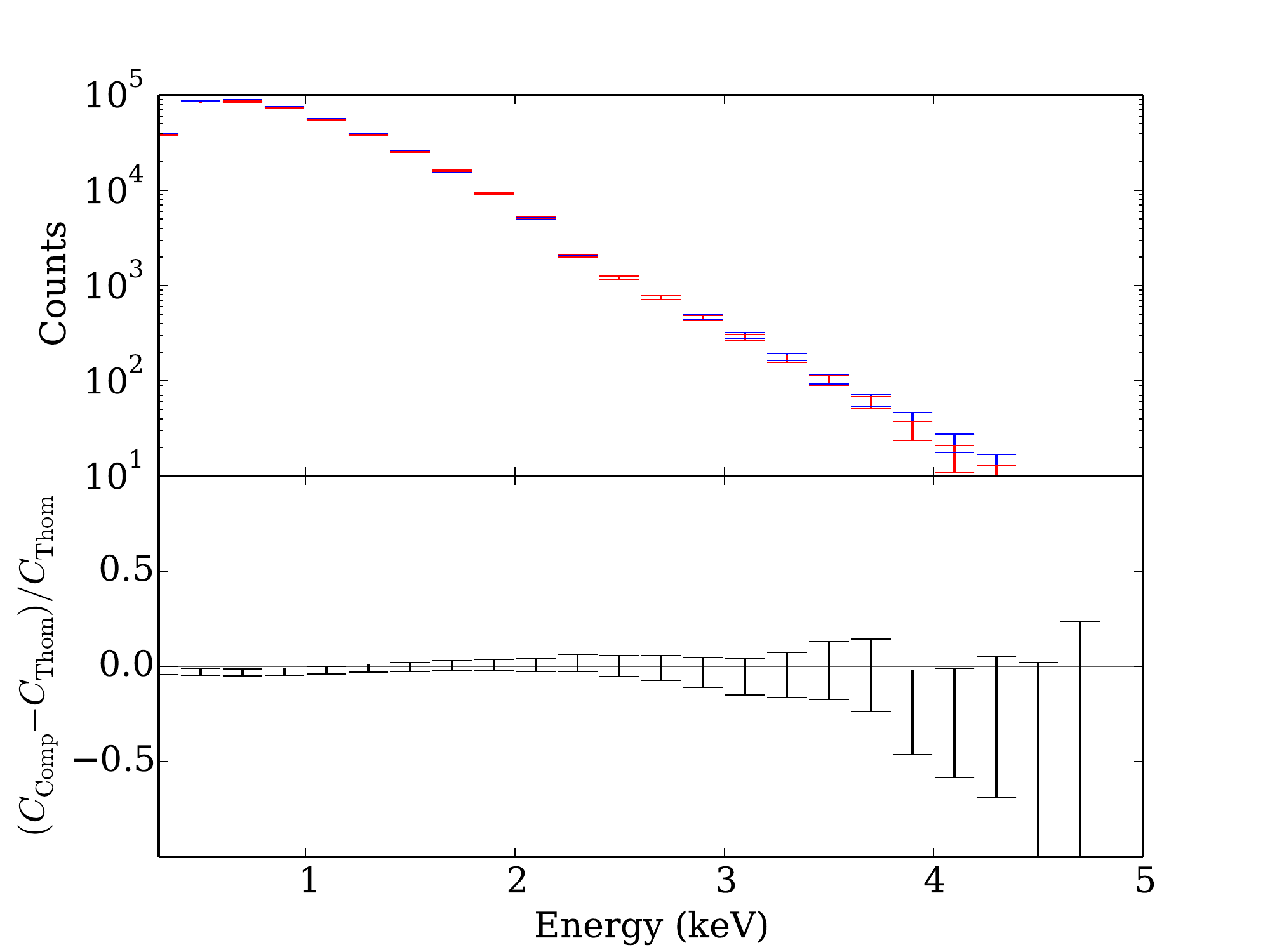}}
\caption{ Upper panel: 
Comparison of two phase-averaged synthetic spectra in terms of  counts detected by NICER with $T =$ 2.0 MK, and one week exposure time of object similar to PSR J0437$-$4715. 
The other parameters of the model are the same as explained in Sect. \ref{sec:methods} (e.g. the spot size remains $5.0 ^\circ$).
The blue bars are calculated with the Thomson model \textsc{McPHAC} and the red bars are for our full Compton model. 
Every twenty adjacent NICER energy bins  are combined to one bin. A calibration error of 1\% is assumed. 
Lower panel: 
The relative difference of the counts predicted by the two models.
The error bars correspond to the combined error of the two data points in a given energy bin relative to the observed counts of Thomson model, 
calculated by $\sqrt{(\sigma_{\mathrm{Comp}}^{2}+\sigma_{\mathrm{Thom}}^{2})}/C_{\mathrm{Thom}}$.}
\label{fig:count_spectra_2MK_rho5}
\end{figure}

\begin{figure}
\resizebox{\hsize}{!}{\includegraphics{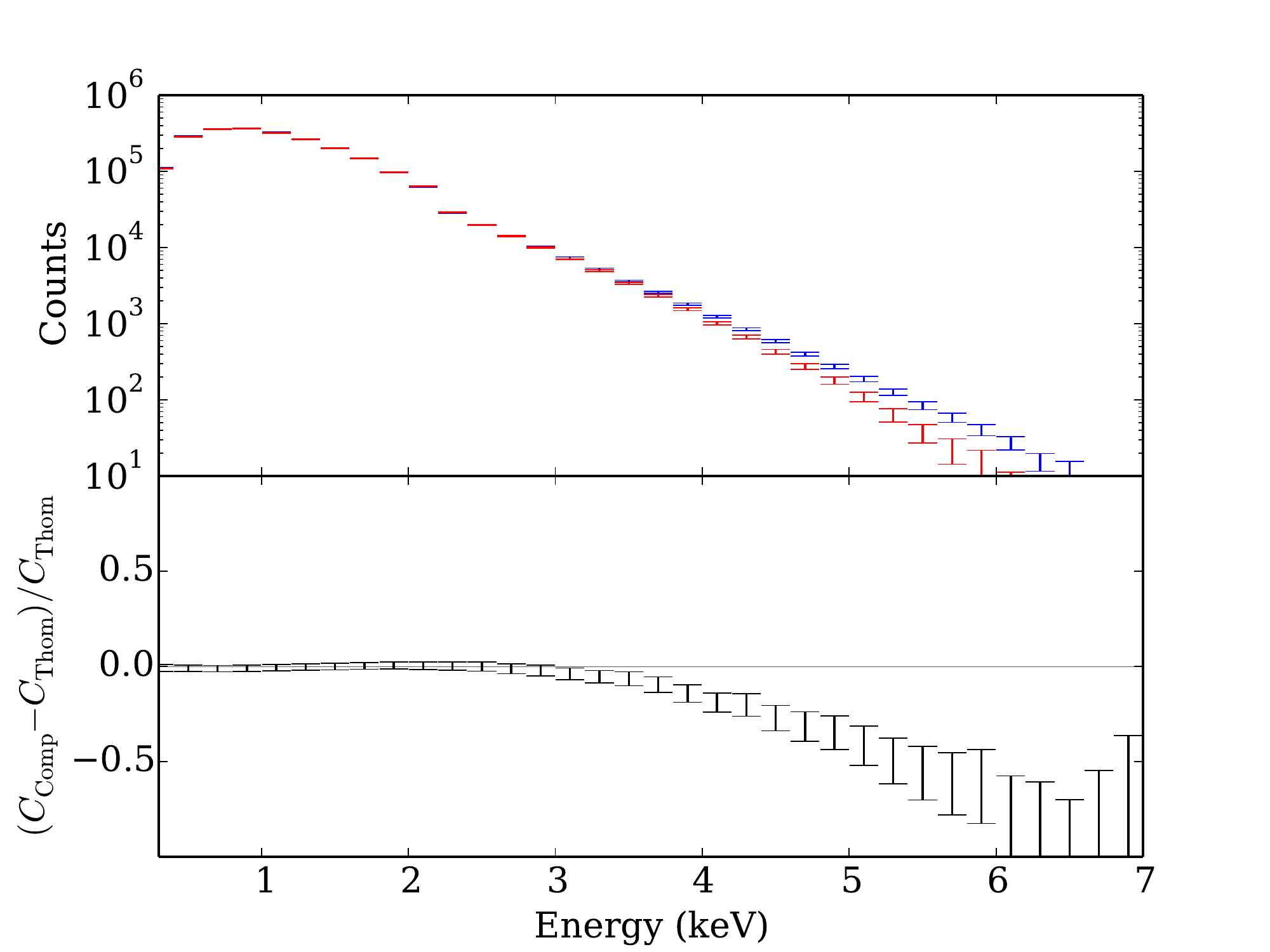}}
\caption{Comparison of two phase-averaged synthetic spectra similarly to Fig.~\ref{fig:count_spectra_2MK_rho5}, but for $T=3.1$ MK.  
The Compton version of the synthetic spectra also shows the spectral part of the synthetic data used in the analysis in Sects. \ref{sec:param_constr_corr} and \ref{sec:param_constr_incorr}.
}
\label{fig:count_spectra_3MK_rho5}
\end{figure}

Taking into account the energy response matrix of the NICER instrument, we also show the modelled phase-averaged count spectra in Fig. \ref{fig:count_spectra_2MK_rho5} with $T=$ 2.0 MK and in Fig. \ref{fig:count_spectra_3MK_rho5} with $T=$ 3.1 MK (other parameters being the same as in Sect. \ref{sec:methods}).
The data produced with the Compton model in Fig. \ref{fig:count_spectra_3MK_rho5} also represent 
our synthetic data in the following sections.
From the figures, we see that the discrepancy between the models at the highest energies can be partly hidden because of only a few detected counts, 
and therefore large statistical errors.
We also assumed the calibration error of the instrument to be 1\%.
In any case, a clearly observable difference above 3 keV remains when $T=$ 3.1 MK.

\begin{figure}
\centering
\includegraphics[width=8cm]{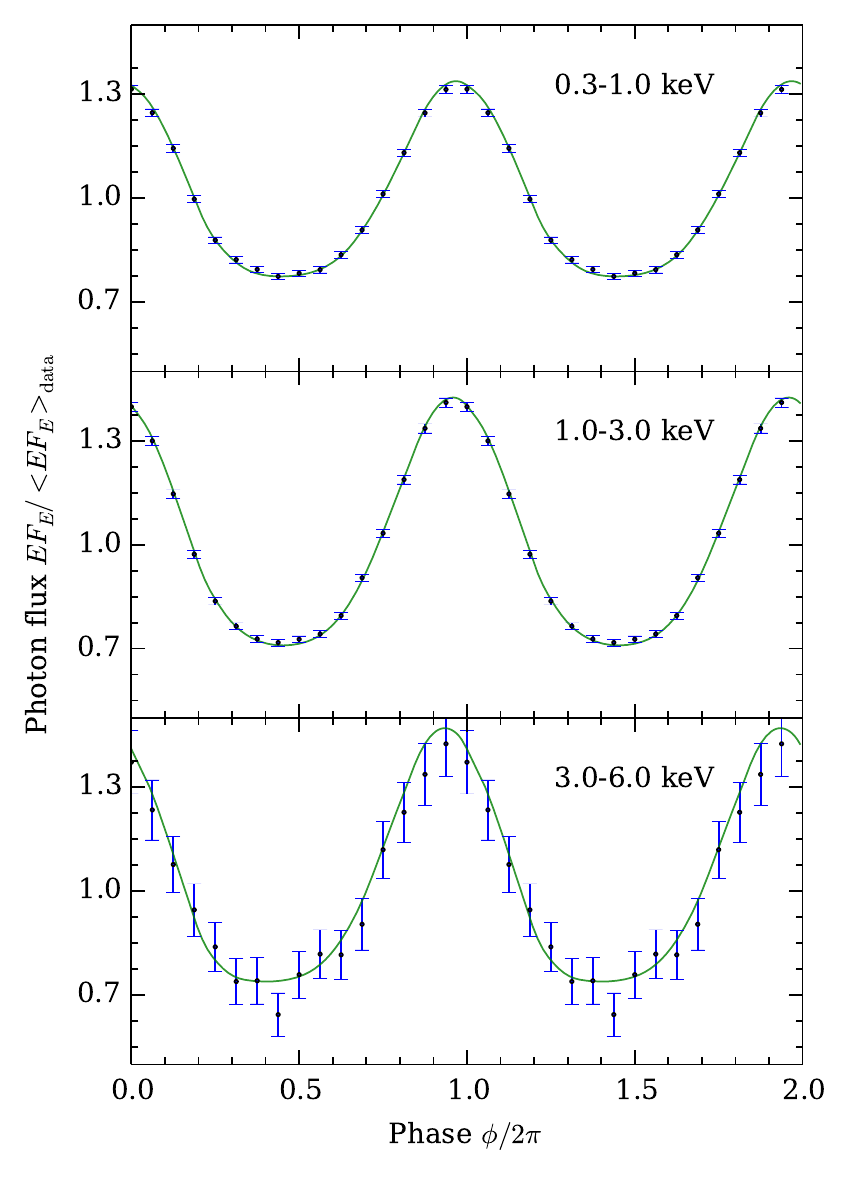} 
\caption{Normalised pulse profiles for the synthetic data simulated using full Compton model. 
For illustration the data are re-binned to 3 energy bins. 
The green solid-line  shows  the  best-fit  solution.  
The contours for posterior density credible regions are not shown as they are very precise and would overlap the line of the best-fit solution. 
The  synthetic  data  converted  to  the physical units using the best-fit model are shown with blue circles, with the error bars shown according to the Poisson noise. 
The assumed calibration error of 1\% is not shown in the error bars.}
\label{fig:pulse_fit_atm_atm}
\end{figure}

\begin{figure}
\centering
\includegraphics[width=9.5cm]{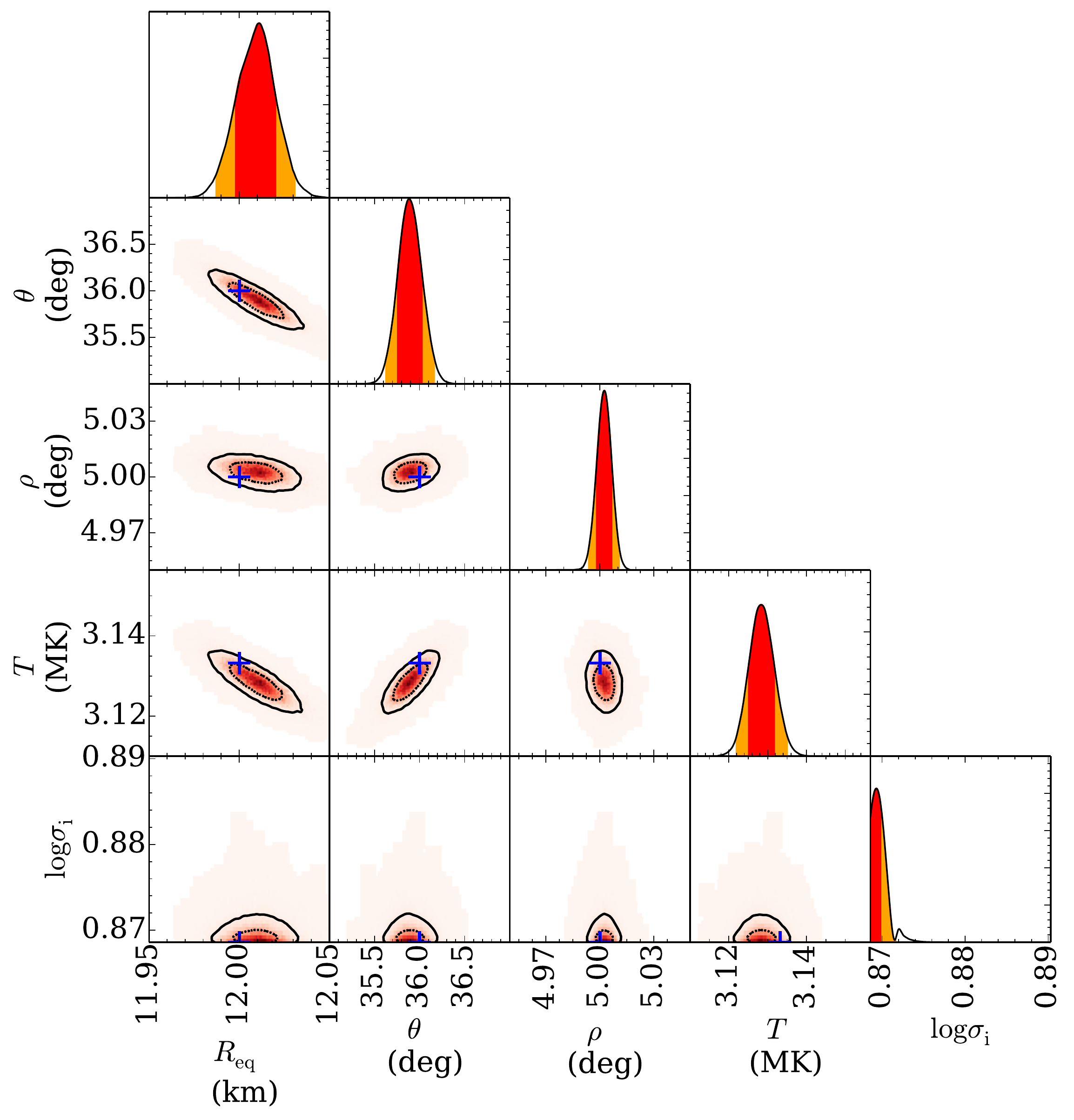} 
\caption{Posterior probability distributions for Markov chain Monte
Carlo 
runs for fitting synthetic data with full Compton model. 
The red colour shows a 68\% and the orange colour a 95\% highest posterior density credible interval. 
In the two-dimensional 
posterior distributions the dashed  contour shows a 68\% and the  solid contour a 95\% highest posterior density credible region. 
The blue crosses show the input value. 
} 
\label{fig:posterior_atm_atm}
\end{figure}

\subsection{Parameter constraints with the correct model}\label{sec:param_constr_corr}

We applied the method described in Sect.~\ref{sec:methods} with full treatment of Compton scattering in the atmosphere, to fit the synthetic data created using the same model.
Due to the ignorance of heating by magnetospheric return currents, our data do not resemble what is expected in real sources.
We still confirm the robustness of our method by getting no strong biases in the constraints for radius and other parameters.
The fitted pulse profiles are shown in Fig.~\ref{fig:pulse_fit_atm_atm} (integrated to three energy bins), and the posterior probability distributions are shown in Fig. ~\ref{fig:posterior_atm_atm}.
The credible limits of all parameters are also listed in Table~\ref{table:conflimits}.
The best-fit solution presented in  Fig.~\ref{fig:pulse_fit_atm_atm} has $\chi^{2}/\mathrm{d.o.f.} = 6711/(6736-6) \approx 1.00$ (for six free parameters including the phase shift), when ignoring the calibration error, which is used only for fitting purposes but not actually present in the synthetic data.

\begin{table*}
  \caption{Most probable values and 68\% and 95\% credible limits for Compton and Thomson models applied to synthetic data.}
\label{table:conflimits}
\centering
  \begin{tabular}[c]{l c c c c c}
    \hline\hline
      Quantity & 95\% lower limit & 68\% lower limit & Most probable value& 68\% upper limit & 95\% upper limit \\ \hline   

     \multicolumn{6}{c}{Compton fit to Compton model} \\ 


      $R_{\mathrm{eq}}$ (km) & $11.99$ & $12.00$ & $12.01$ & $12.02$ & $12.03$  \\ 
      $\theta$ ($\deg$) & $35.6$ & $35.8$ & $35.9$ & $36.0$ & $36.2$ \\ 
      $\rho$ ($\deg$) & $4.99$ & $5.00$ & $5.00$ & $5.01$ & $5.01$ \\ 
      $T$ (MK) & $3.121$ & $3.125$ & $3.128$ & $3.131$ & $3.135$ \\ 
      $\log(\sigma_{\mathrm{i}})$ & $0.868$ & $0.869$ & $0.869$ & $0.870$ & $0.872$ \\ \hline

      \multicolumn{6}{c}{\textsc{McPHAC} fit to Compton model} \\ 


      $R_{\mathrm{eq}}$ (km) & $11.99$ & $12.00$ & $12.01$ & $12.02$ & $12.03$  \\ 
      $\theta$ ($\deg$) & $35.5$ & $35.6$ & $35.8$ & $35.9$ & $36.0$ \\ 
      $\rho$ ($\deg$) & $4.95$ & $4.96$ & $4.96$ & $4.96$ & $4.97$ \\ 
      $T$ (MK) & $3.136$ & $3.139$ & $3.142$ & $3.145$ & $3.148$ \\ 
      $\log(\sigma_{\mathrm{i}})$ & $0.980$ & $0.993$ & $1.007$ & $1.020$ & $1.032$ \\ \hline

  \end{tabular}
\tablefoot{The quantities shown in the Table are equatorial radius $R_{\mathrm{eq}}$, spot co-latitude $\theta$, spot angular radius $\rho$,   hotspot temperature $T$, and intrinsic scatter $\log \sigma_{\mathrm{i}}$.
The correct values for the model parameters are $R_{\mathrm{eq}} = 12$ km, $\theta = 36^\circ$, $\rho = 5^\circ$, and $T = 3.133$ MK.}  
\end{table*}

As expected, the model accurately 
describes the synthetic data, as seen in the posterior probability distribution for intrinsic scatter $\sigma_{\mathrm{i}}$.
The mean $\log \sigma_{\mathrm{i}} <  1$ of the posterior translates to an error of less than ten counts in each phase-energy bin. This effectively means zero intrinsic scatter, as it is significantly smaller than the Poisson noise of the data, which is $56$ counts on average in each fitted phase-energy bin. 
For the radius, we find the 68\%  (95\%) limits and the most probable value as $R_{\mathrm{eq}} = 12.01^{+0.01 ~(0.02)}_{-0.01 ~(0.02)}$ km.
We note that this, and the other limits presented here and in the following section, are considerably tighter than what is expected, if comparing, for example, to the approximation in Eq. (5) by \citet{POC14}, which has been used to predict 5\% accuracy for the NICER targets.
With our model parameters and the amount of detected counts ($4\times 10^{7}$), we should have about 1\% accuracy.
Our even tighter limits could be due to the anisotropic effects (ignored in the aforementioned equation), which can strongly increase the second harmonic of the pulse profile signal \citep{PB06}, and thus decrease the uncertainty in its measurement (as we regard the atmospheric effects to be known).
In any case, this is not critical given that we are only interested in the differences between the two spectral models.
Similarly, tight constraints are found for other parameters so that the correct point remains  
inside their 68\% limits, except for the temperature where the correct point is slightly offset towards smaller values, but is still inside the 95\% limits.

\begin{figure}
\centering
\includegraphics[width=8cm]{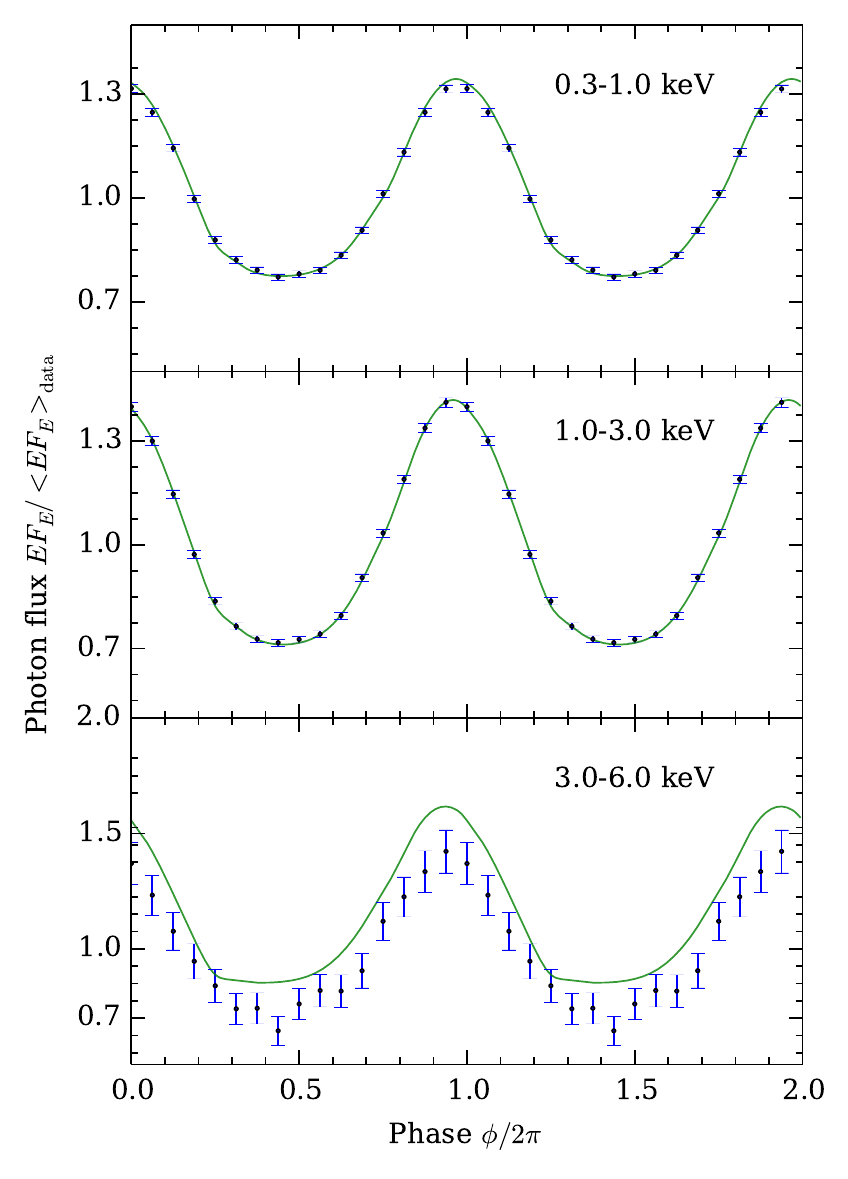} 
\caption{
Normalised pulse profiles for the synthetic data simulated using \textsc{McPHAC} Thomson model. Identifying information here is the same as in Fig.~\ref{fig:pulse_fit_atm_atm}.}
\label{fig:pulse_fit_mcp_atm}
\end{figure}

\begin{figure}
\centering
\includegraphics[width=8.8cm]{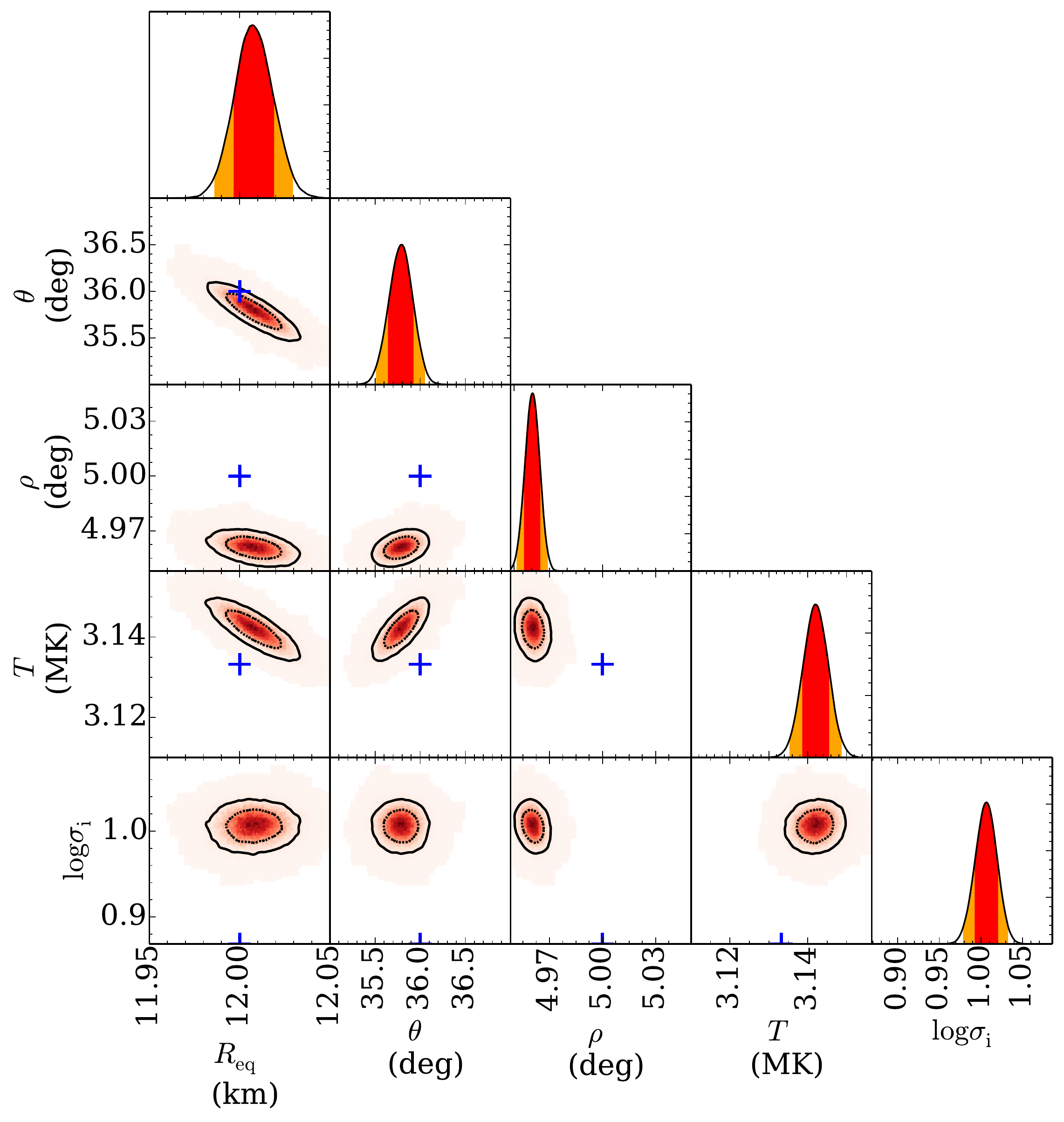} 
\caption{
Posterior probability distributions for Markov chain Monte Carlo runs for fitting synthetic data with \textsc{McPHAC} Thomson model. Identifying information here is the same as in Fig.~\ref{fig:posterior_atm_atm}.
}
\label{fig:posterior_mcp_atm}
\end{figure}

\subsection{Parameter constraints with the incorrect model}\label{sec:param_constr_incorr}

We also applied the method described in Sect.~\ref{sec:methods}, using NS atmosphere model \textsc{McPHAC} to fit the synthetic data that were created using the full Compton scattering model.
The fitted pulse profiles, integrated to three energy bins, are shown in Fig.~\ref{fig:pulse_fit_mcp_atm} for illustration.
We see that the fits are worse at the highest energies due to a large difference in the spectral shapes.
The posterior probability distributions are shown in Fig.~\ref{fig:posterior_mcp_atm}, and the credible limits are listed in Table \ref{table:conflimits}.

We find that the constraints for radius are still not biased, but very close to those obtained in the previous section, since $R_{\mathrm{eq}} = 12.01^{+0.01 ~(0.02)}_{-0.01 ~(0.02)}$ km.
However, the credible limits for the temperature, and especially for the size of the spot, are clearly different: the temperature is higher while the spot size is smaller 
than the correct values.
Neither of them agrees with the $95$\% limits. 
The best-fit solution presented in  Fig. \ref{fig:pulse_fit_mcp_atm} has $\chi^{2}/\mathrm{d.o.f.} = 11142/6730 \approx 1.66$.
In addition, according to the notably higher intrinsic scatter $\sigma_{\mathrm{i}}$, the model does not describe the synthetic data as well as the correct model that includes Compton scattering (although $\sigma_{\mathrm{i}}$ is still effectively very small compared to the Poisson noise).

We also calculated the results assuming two other NS masses.
With otherwise a similar setup as discussed above, the synthetic data were created with  masses $1.4 \msun$ and $2.0 \msun$ and then fitted with both Thomson and Compton models. 
We find no major difference compared to the already presented results using an NS mass of $1.76 \msun$.
Although, in the case of the Thomson model, 
biases in temperature $T$ and spot size $\rho$ are found to depend on the NS mass.
The bias is always significant and 
tends towards the same direction, but it is higher with higher masses.
Further, the slope between radius $R_{\mathrm{eq}}$ and magnetic co-latitude $\theta$ in the two-dimensional posterior probability histogram is different for every mass  because the star is more oblate with higher masses. 
In all cases the input radius is still obtained at least within the 95\% limits.

\section{Conclusions}\label{sec:conclusions}

We studied the possible outcomes of using Thomson scattering approximation in the atmosphere calculation instead of a full Compton scattering model when trying to constrain NS parameters from RMP pulse profile observations of NICER.
Our spectral comparisons showed that the difference in the observed spectrum may not be detected, due to the low count rate at highest energies, if the temperature of the emitting hotspot is $T=2$ MK.
However, in case of $T \approx 3$ MK, a significant discrepancy can be observed.

We simulated and fitted synthetic data, based on the Compton atmosphere with $T=3.1$ MK and the NICER target PSR J0437$-$4715, which is expected to give some of the most constraining limits to NS radius.
Fitting with the same Compton model, we obtained very tight limits for the NS parameters without strong biases,  demonstrating the robustness of our method. 
Likewise, fitting with the Thomson model resulted in very similar constraints on the radius.
However, the obtained size and the temperature of the hotspot were significantly different.
The exact credible limits should not be taken too seriously, as we have exaggerated the predicted count rate in order to emphasize the differences between the two spectral models.

According to our results, Compton scattering seems to be unimportant in obtaining accurate radius constraints for RMPs, at least in the case of a similar model and comparable data to that used here.
However, for the interpretation of the data from a mission that is more sensitive at high energies and observes more counts at the energies around and above 3 keV, the effects of Compton scattering would need to be taken into account in a precise manner. 
These effects will be even more important for atmospheres heated in the surface layers by bombarding particles, 
which were not considered in this paper but are expected to be present in real sources.
 
\section*{Acknowledgments}
This research was supported by the University of Turku Graduate School in Physical and Chemical Sciences (TS), by the  Ministry of Science and Higher Education of the Russian Federation grant 14.W03.31.0021 (JP, VFS), the Deutsche Forschungsgemeinschaft (DFG) grant WE 1312/51-1, the travel grant of the German Academic Exchange Service (DAAD, project 57405000), and the Academy of Finland travel grant 317552 (JP).
The computer resources of the Finnish IT Center for Science (CSC) and the Finnish Grid and Cloud Infrastructure project are acknowledged.





\end{document}